\documentclass[iop]{emulateapj}
\usepackage{epsfig}
\usepackage{subfigure}
\usepackage{multirow}

\newcommand{\half}{{\textstyle\frac{1}{2}}}
\newcommand\an{{\em Astron.\ Nachr.}}

\shorttitle{Coupling strengths of mixed modes in subgiants}
\shortauthors{Benomar et al.}

\begin{document}


\title{Masses of subgiant stars from asteroseismology using the coupling strengths of mixed modes}


\author{O. Benomar,\altaffilmark{1}
T.R.~Bedding,\altaffilmark{1}
D.~Stello,\altaffilmark{1} 
S.~Deheuvels,\altaffilmark{2} 
T.R.~White\altaffilmark{1} and
J.~Christensen-Dalsgaard\altaffilmark{3}}

\altaffiltext{1}{Sydney Institute for Astronomy (SIfA), School of Physics,
  University of Sydney, NSW 2006, Australia} 
\altaffiltext{2}{Department of Astronomy, Yale University, P.O. Box 208101, New Haven, CT 06520-8101, USA}
\altaffiltext{3}{{Danish AsteroSeismology Centre (DASC), Department of Physics and Astronomy, Aarhus University, DK-8000 Aarhus C, Denmark}}

\begin{abstract}
Since few decades, asteroseismology, the study of stellar oscillations, enables us to probe
the interiors of stars with great precision. It allows stringent tests of
stellar models and can provide accurate radii, masses and ages for
individual stars. Of particular interest are the mixed modes that occur in subgiant solar-like stars since
they can place very strong constraints on stellar ages. Here we
measure the characteristics of the mixed modes, particularly the coupling
strength, using a grid of stellar models for stars with masses between
$0.9$ and $1.5$ $M_\odot$.  We show that the coupling strength of the
$\ell=1$ mixed modes is predominantly a function of stellar mass and appears to be
independent of metallicity. This should allow an accurate mass evaluation, further increasing the usefulness of mixed modes in subgiants as asteroseismic tools.
\end{abstract}


\section{Introduction}
	\label{sec:intro}

Asteroseismology allows stringent tests of
stellar models and can provide accurate fundamental properties of individual stars. For solar-type stars on the main sequence, the observed
oscillations are $p$~modes, for which the restoring force arises from the
pressure gradient.  These are approximately regularly spaced in frequency,
following closely the so-called asymptotic relation
\citep{Vandakurov1967AZh,tassoul1980,gough1986}.  However, the oscillations of
post-main-sequence stars show departures from this regularity that are due
to the presence of mixed modes.

Mixed modes have $p$-mode character in the stellar envelope and $g$-mode
character in the core. %($g$~modes are gravity modes, for which the restoring force is buoyancy).  
They occur in evolved stars (subgiants and red
giants), in which the large density gradient outside the core effectively
divides the star into two coupled cavities.  This leads to %the phenomenon of 
{\em mode bumping}, in which mode frequencies are shifted from their
regular spacing and no longer follow the asymptotic relation.  Mode bumping
in subgiant stars was first observed and modeled in $\eta$~Boo
\citep{Kjeldsen1995b,kjeldsen2003,christensen1995,carrier2005} and
$\beta$~Hyi \citep{bedding2007,Brandao2011}.  More recently, asteroseismic space missions have produced many more examples, including the
CoRoT target HD~49835 \citep{Deheuvels2010a} and a growing number of {\em
Kepler\/} stars (e.g., \citealt{Metcalfe2010, Mathur2010, Campante2011}).

Mixed modes carry valuable information on the internal structure and
evolutionary state of stars.  Their frequencies change %relatively 
quickly with time as they undergo avoided crossings \citep{Osaki1975,
Aizenman1977}, potentially providing stellar ages with a precision down to a few Myr, %corresponding to
 or a relative uncertainty of $\approx 1\%$
\citep{Metcalfe2010}. \\
Mixed modes arise from a resonant coupling between $p$ and $g$~modes %.
%Resonances are common in many fields of physics 
and can be well represented by a system of coupled oscillators.  In this Letter, we use such a
representation to model their frequencies %of mixed modes 
in subgiant stars and determine %the strength of 
the coupling strength between the modes.  We show that
the coupling strength depends strongly on stellar mass but only weakly (or not at all)
on metallicity, hence lifting the often problematic degeneracy between
those two variables.

 \section{A model for mixed modes}
 \label{sec:gene}

%As discussed above, 
Mixed modes in stars occur when the $p$ and $g$ modes
are coupled, which leads to avoided crossings and mode bumping.
\cite{Deheuvels2010b} suggested that an avoided crossing in a subgiant star
can be well-represented by a system of $(n-1)$ $p$-mode oscillators, each
coupled with a single $g$~mode.  This can be modeled by a system of $n$
differential equations, one for each of the $p$~modes and one for the
$g$~mode:
\begin{eqnarray} \label{eq:LS_system} 
	\frac{d^2 y_1(t)}{dt^2} &=& - \omega_{\pi_1}^2 y_1 + \alpha_{1,n} y_n \nonumber\\
	&\vdots& \\%\nonumber\\
%% 	&.& \\
%% 	&.& \\ \nonumber
	\frac{d^2 y_{n-1}(t)}{dt^2} &=& - \omega_{\pi_{n-1}}^2 y_{n-1} + \alpha_{n-1,n} y_n \nonumber\\
	\frac{d^2 y_{n}(t)}{dt^2} &=& - \omega_{\gamma}^2 y_{n} + \alpha_{1,n} y_1 + ... +  \alpha_{n-1,n} y_{n-1} \nonumber
\end{eqnarray}
The $y_i(t)$ terms are the displacements of the modes and $\alpha_{i,n}$ are the coupling coefficients between the $i$th $p$~mode and the $g$~mode.
The frequencies $\omega_{\pi_i}$ and $\omega_\gamma$ %(equal to $2\pi \nu$) 
correspond to fictitious, pure $p$ and $g$ modes
that would exist if their cavities were not coupled. In order to avoid
ambiguity, we follow \cite{Aizenman1977} by referring to
these as $\pi$ and $\gamma$ modes, respectively (see also \citealt{bedding2011a}).

We seek oscillatory solutions of Eq.\,\ref{eq:LS_system} with angular
frequencies $\Omega=\left\{\omega_1, \omega_2, ..., \omega_n \right\}$, which means we need to solve the following system:
\begin{equation} \label{eq:LS}
\mbox{\boldmath$A$} Y=\Omega^2 Y,
\end{equation} 
where $Y=\left\{ y_1, y_2, ..., y_n \right\}$ and
\begin{equation} \label{eq:mat_system}
\mbox{\boldmath$A$}= \left( \begin{array}{ccccc}
\omega_{\pi_1}^2 & \cdots     &    0           & -\alpha \\
 \vdots    & \ddots     &    0           & -\alpha\\
    0      &  \cdots    & \omega_{\pi_{n-1}}^2 & -\alpha \\
   -\alpha &  \cdots    & -\alpha        & \omega_\gamma^2
\end{array} \right).
\end{equation}
Following \cite{Deheuvels2010b}, we assumed the coupling
$\alpha_{i,n} \equiv \alpha$ to be the same between all modes.  This is a
reasonable assumption provided $\alpha$ varies slowly with frequency, which
turns out to be the case in the models we have studied.

%Note that 
In more evolved stars (red giants), the situation is
reversed and several $g$-modes may be coupled to a single $p$-mode
\citep{Dupret2009, christensen2011b, bedding2011a, Stello2011}. Between these two
extremes, one can consider intermediate cases with multiple
$g$-modes and $p$-modes \citep[e.g.,][]{DiMauro2011}, whose power spectra
may become hard to interpret.

\begin{figure}
\begin{center}
\includegraphics*[angle=90,height=6.5cm]{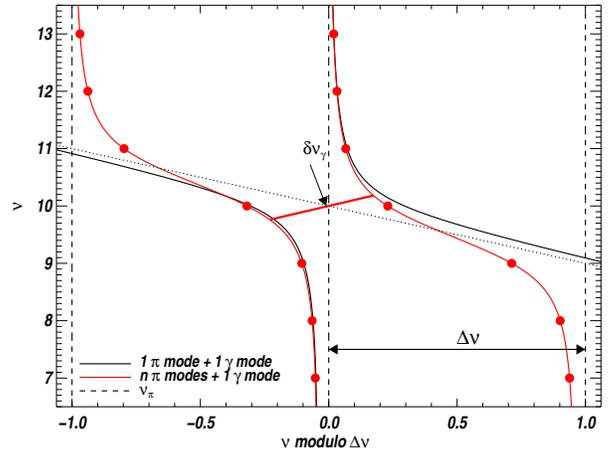}  
\caption{ Replicated \'echelle diagram showing the effect of the coupling
between the $\pi$ and $\gamma$ modes in two configurations: two coupled
modes (black), and several coupled modes (red) with $\pi$-modes frequencies spaced by $\Delta\nu$. $\nu_\gamma$ is arbitrarily fixed to $10$. Where the $\pi$ and $\gamma$ modes cross, the frequency separation
%has a value of 
is $\delta\nu_\gamma$ (diagonal red line). $\delta\nu_\gamma$ is %This distance is
almost independent of $n$, the number of interacting modes.}
\label{fig:minimmal_distance}
\end{center}
\end{figure}

For the case of $n=2$, which is a single $\pi$~mode coupling to a
$\gamma$~mode, Eq.\,\ref{eq:LS} can be solved analytically to give a pair of
solutions \citep{Deheuvels2010b}:
\begin{equation} \label{eq:solutions_two_modes}
	\omega_\pm^2 = \frac{\omega_\pi^2 + \omega_\gamma^2 }{2} \pm \half
      	               \sqrt{ (\omega_\pi^2 - \omega_\gamma^2)^2 + 4\alpha^2}.
\end{equation}
The black curves in Figure\,\ref{fig:minimmal_distance} shows these two solutions
in a replicated \'echelle diagram \citep{bedding2011b}.

We note that the coefficient $\alpha$ is not an ideal measure of the
coupling strength between $\pi$ and $\gamma$ modes because it varies with
frequency and is therefore different for each avoided crossing.  A better
measure is the frequency separation of the two solutions at
the avoided crossing.  That is, we focus on the point of resonance by
setting $\omega_\gamma = \omega_\pi$ and measure the frequency difference
between the two solutions:
\begin{equation}
	\delta\omega_\gamma = \omega_+ - \omega_-.
\end{equation}
It follows from Eq.\,\ref{eq:solutions_two_modes}
that
\begin{equation}
	\delta\omega_\gamma = \alpha / \omega_\gamma, \label{eq:minimal_distance}
\end{equation}
or, in cyclic frequency,
\begin{equation} \label{eq:delta_nu_gamma}
\delta\nu_\gamma = \delta\omega_\gamma / 2 \pi = \frac{\alpha}{4 \pi^2 \nu_\gamma}.
\end{equation}

For the case of $n>2$, which involves several $\pi$ modes coupling to a
$\gamma$ mode, Eq.\,\ref{eq:LS} can be solved numerically. %by inverting the
%matrix $\mbox{\boldmath$A$}$.  
 We find a set of solutions, %two of them are
shown by the red curves in Figure\,\ref{fig:minimmal_distance} for the case of
$n=8$. The value of $\delta\nu_\gamma$ is shown by the
solid red line. \\% in Figure\,\ref{fig:minimmal_distance}.  
The minimal separation is almost independent of the number of interacting modes (black
and red lines are close to each other at $\omega_\pi \simeq
\omega_\gamma$). The stronger the coupling, the larger the separation. 
%Note that since $\delta\nu_\gamma \propto \frac{1}{\nu_\gamma}$, the line drawn in
%Figure\,\ref{fig:minimmal_distance} does not appear to be the shortest in a
%plot of $\nu$ vs.\ $\nu \bmod \Delta\nu$.

\section{Fitting method}

For each avoided crossing we have a sequence of observed frequencies that
correspond to the mixed modes.  Each observed frequency must satisfy %the linear system in 
Eq.\,\ref{eq:LS}.  Provided we have enough
observed frequencies, we can determine the elements of the
matrix~$\mbox{\boldmath$A$}$, which gives %the characteristics of the avoided crossing (
the $\pi$ and $\gamma$ mode frequencies, and the coupling strength~$\alpha$.

To carry out this fitting process robustly, we apply a Bayesian approach
(Maximum a Posteriori Approach, hereafter MAP).  Assuming a likelihood
function $L(\nu_{\rm obs}|\mbox{\boldmath$A$} )$, one can regularize this
function by penalizing it with a quantity $p(\mbox{\boldmath$A$})$, called
the prior, where $\nu_{\rm obs}$ represents the observed frequencies
(the data).
Using Bayes' theorem, we write the statistical criteria
defining the fit quality that we seek to maximize as
\begin{equation}
	\ln[ p(\mbox{\boldmath$A$} |\nu_{\rm obs}) ] = \ln [ L(\nu_{\rm
	obs}|\mbox{\boldmath$A$}) ] + \ln[ p(\mbox{\boldmath$A$})] -
	\ln[C].
\end{equation}
Here, $C$ is a normalization constant and $p(\mbox{\boldmath$A$} |\nu_{\rm
obs})$ is the posterior probability (the probability of
$\mbox{\boldmath$A$}$ given ~$\nu_{\rm obs}$).

The linear system in Eq.\,\ref{eq:LS} must be solved at each iteration of
the maximization process in order to compare the observed frequencies
$\nu_{\rm obs}$ with the calculated frequencies $\nu_{\rm calc}$. The expression we
choose to minimize is $\half \sum_i \left(\nu_{\rm obs}(n) - \nu_{\rm
calc}(n)\right)^2/\sigma^2_n$. Thus the log-likelihood is, %can be written,
\begin{equation}
	\ln [ L(\nu_{\rm obs}|\mbox{\boldmath$A$}) ] \propto - \half \sum_n
	(\nu_{\rm obs}(n) - \nu_{\rm calc}(n))^2/\sigma^2_n. 
\end{equation}
where $\sigma_n=1$ for model frequencies.

Having defined the likelihood, we must choose expressions for the
priors.  No priors on $\alpha$ and $\nu_\gamma$ were applied and our
attention will focus on ~$\nu_{\pi_i}$.  The pseudo-modes $\pi$ and
$\gamma$ are expected to behave as pure $p$ and $g$ modes. To a good
approximation, the $\pi$-modes should follow the asymptotic relation
\citep{tassoul1980}:
\begin{equation} \label{eq:ass_law}
	\nu_\pi(n) = ( n + \frac{\ell}{2} + \epsilon) \Delta\nu - \delta\nu_{0\ell},
\end{equation}
where $n$ is the radial order, $\epsilon$ is an offset associated with
stellar surface effects, $\Delta\nu$ is the large separation (related to
the mean density of the star), and $\delta\nu_{0\ell}$ is sensitive to the core
properties.  In this Letter, we only consider the case of $\ell=1$ (dipole
modes).

Discontinuities inside stars introduce frequency oscillations as a function
of $n$, %which are 
not included in the asymptotic relation. %However, 
 These oscillations are not expected to be significant, say less than a few
percent of $\Delta\nu$. Thus the $\pi$-mode frequency variations are
expected to be smooth, as seen in the \'echelle diagram for the $p$-modes
of main sequence stars.  A way to satisfy this condition is to impose a
smoothness condition on some $p$th derivative terms of $\nu_\pi(n)$. We note that the second derivative in $n$ of Eq.\,\ref{eq:ass_law},
denoted $\Delta^2\nu (n)$, leads to $\Delta^2\nu (n)=\frac{\partial^2 \nu_{\pi}}{ \partial
n^2} \approx 0$. Thus one may limit local strong deviations from a regular
pattern by imposing a Gaussian prior on $\Delta^2\nu (n)$,
\begin{equation}
	p( \Delta^2\nu (n) ) = \frac{1}{\sqrt{2 \pi} \sigma_{\Delta^2\nu}} \exp\left[- \half \left(\frac{\Delta^2\nu (n)}{\sigma_{\Delta^2\nu}} \right)^2 \right].
\end{equation}
Here, $\sigma_{\Delta^2\nu}$ plays the role of a relaxation constraint and must
be chosen to ensure enough freedom, but not too much, in order to
efficiently smooth the frequency profile.  A trial-and-error procedure
showed that $\sigma_{\Delta^2\nu} \approx 2$ $\mu$Hz offers a good compromise.
%for both models and observations.  
With such a smoothness condition, the
$\pi$-mode deviation from a strictly regular pattern of frequencies is
locally described by a second-order polynomial function of $n$, and the
solution belongs to the family of spline functions. The smoothness condition acts locally and does not restrict the $\pi$-modes
to follow the asymptotic relation, globally.

We define an additional (global) condition, in order to avoid strong departures from Eq.\ref{eq:ass_law}.
%the asymptotic relation (Eq.\ref{eq:ass_law}). 
We expect the $\ell=0$ modes and $\ell=1$ $\pi$ modes to be
distributed along two parallel ridges in the \'echelle diagram. %In other words
Hence, the large separation $\Delta\nu$ of the $\ell=0$ mode%, hereafter referred to as
, is expected to be approximately equal to the large
separation $\Delta\nu_\pi$ of the $\ell=1$ $\pi$ modes. %, hereafter referred to as
Thus a second Gaussian prior was applied to the quantity $\Delta\nu_\pi - \Delta\nu$, with $\sigma_{\Delta\nu_\pi} \approx 1/\sqrt{N}$ $\mu$Hz, $N$ being the number of modes.%on the mean
%value of the first derivative of $\nu_\pi(n)$, $\Delta\nu_\pi \equiv \left\langle \frac{ \partial \nu_\pi(n) }{\partial n}
%\right\rangle $.%,
%\begin{equation}
%	p( \Delta\nu_\pi) = \frac{1}{\sqrt{2 \pi} \sigma_{\Delta\nu_\pi}}\exp\left[- \half \left(\frac{\Delta\nu_\pi - \Delta\nu }{\sigma_{\Delta\nu_\pi}} \right)^2 \right].
%\end{equation} 
%A trial-and-error procedure showed that $\sigma_{\Delta\nu_\pi} \approx 1/\sqrt{N}$ $\mu$Hz, with $N$ the number of modes, offers a good compromise.
 
\section{Results for stellar models} \label{sec:models}

We have applied our method to frequencies of $\ell=1$ modes calculated from
stellar models.  Models were selected from the grid described in detail
by \cite{Stello2009}.  This grid was generated with the ASTEC code
\citep{JCD2008ASTEC} using the simple but fast EFF equation of state
\citep{Eggleton1973}, a fixed mixed length parameter at $\alpha_{\rm
MLT}=1.8$, and an initial hydrogen abundance of $X=0.7$. The opacities were
calculated using the solar mixture \citep{Grevesse1993} and the opacity
tables of \cite{rogers1995a} and \cite{kurucz1991}. Rotation, overshooting
and diffusion were not included.
	
Models with masses in the range $0.9$ to $1.5\,M_\odot$ were explored with
a step size of $0.1\,M_\odot$. All models were computed for solar
metallicity ($Z=0.017$).  In addition, lower and higher metallicities
($Z=0.011$ and $Z=0.028$) were considered for a subset of masses
($1.0\,M_\odot$, $1.2\,M_\odot$, $1.3\,M_\odot$ and $1.5\,M_\odot$).

\begin{figure}
\begin{center}
\includegraphics*[angle=90,height=6.5cm]{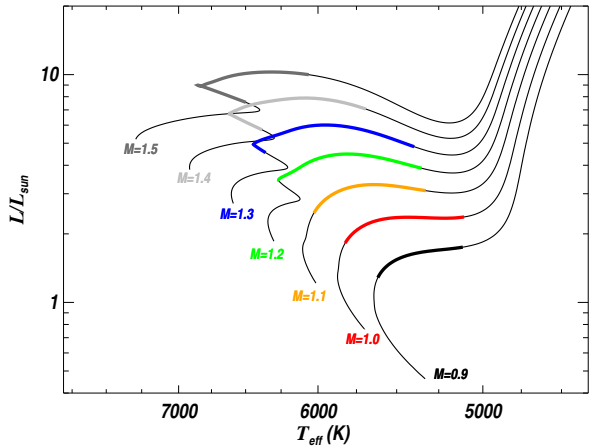} 
\caption{HR diagram for models with $M=0.9 - 1.5$M$_\odot$ at solar
  metallicity. The thick colored lines show the range over which the three
  highest frequency avoided crossings were tracked.} 
\label{fig:HR_diagram}
\end{center}
\end{figure}

In general, stars have mixed $\ell=1$ modes while still on the main
sequence.  However, these occur at low frequencies that lie outside the
envelope of observable modes.  Mode bumping only starts to be detectable
once a star has entered the subgiant phase.  As discussed in the
Introduction, each avoided crossing is associated with a g~mode trapped in
the core of the star (a~$\gamma$~mode; see also
\citealt{deheuvels+michel2010-mu-gradient, bedding2011a}).  %The frequencies of the 
The $\gamma$-mode frequencies increase as the star evolves and we tracked each one.
As additional avoided crossings appeared at low frequency, they were also
incorporated into Eq.\,\ref{eq:LS}. In this way, the stellar model was followed
during a significant part of the subgiant phase, as shown in the HR diagram
in Figure\,\ref{fig:HR_diagram}.  Towards the end of this phase, the frequency
patterns became too complex (too few modes between each avoided crossing)
and a stable fit was difficult to obtain. We secured reliable results for
models with up to three observable avoided crossings, each of them %providing the two parameters that interest us for the present study, namely
characterized by its frequency $\nu_\gamma$, and its coupling strength (measured by~$\delta\nu_\gamma$
--- see Eq.\,\ref{eq:delta_nu_gamma}).

\begin{figure}
\begin{center}
\includegraphics*[angle=90,height=6.3cm]{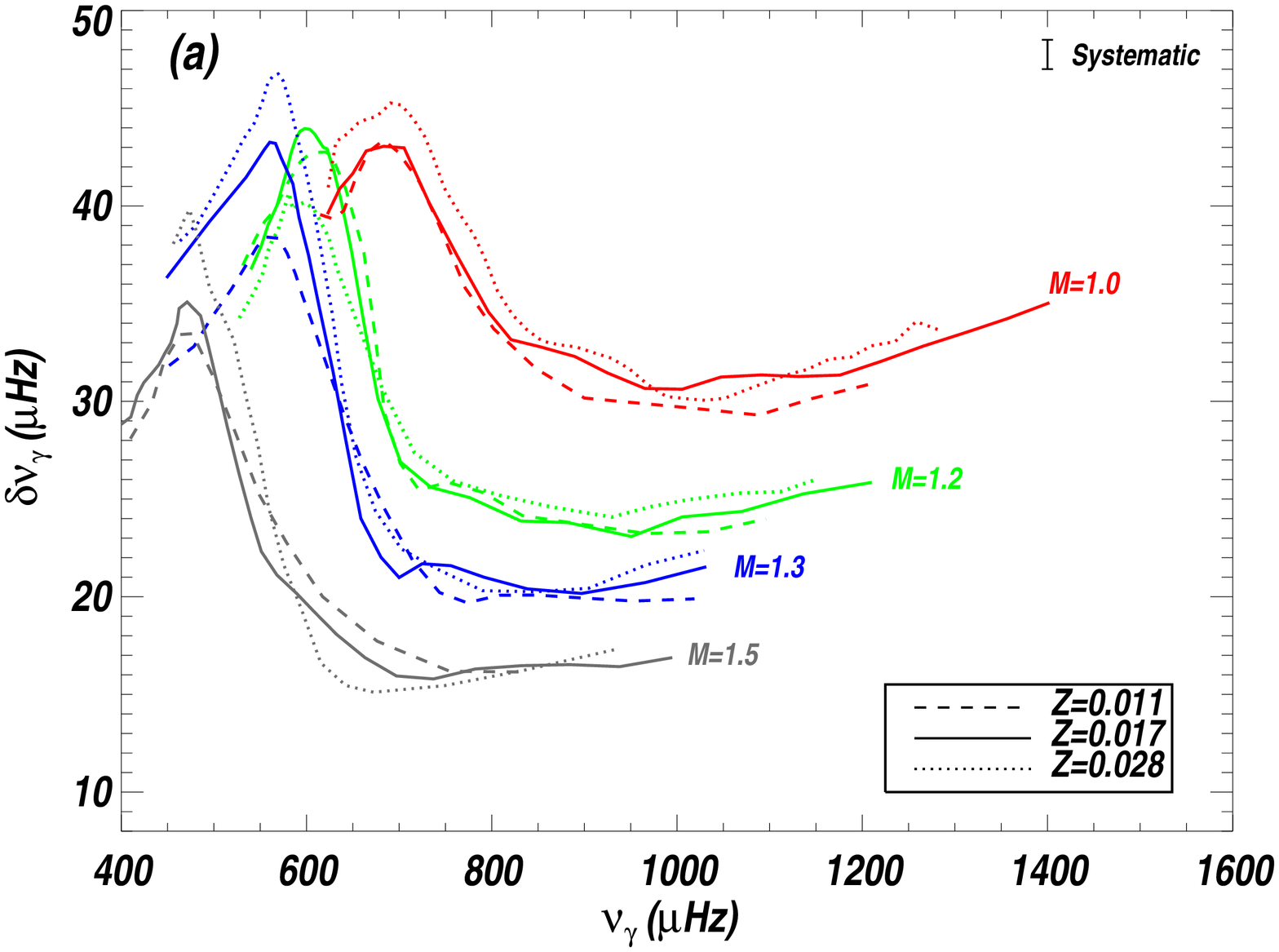} 
\includegraphics*[angle=90,height=6.3cm]{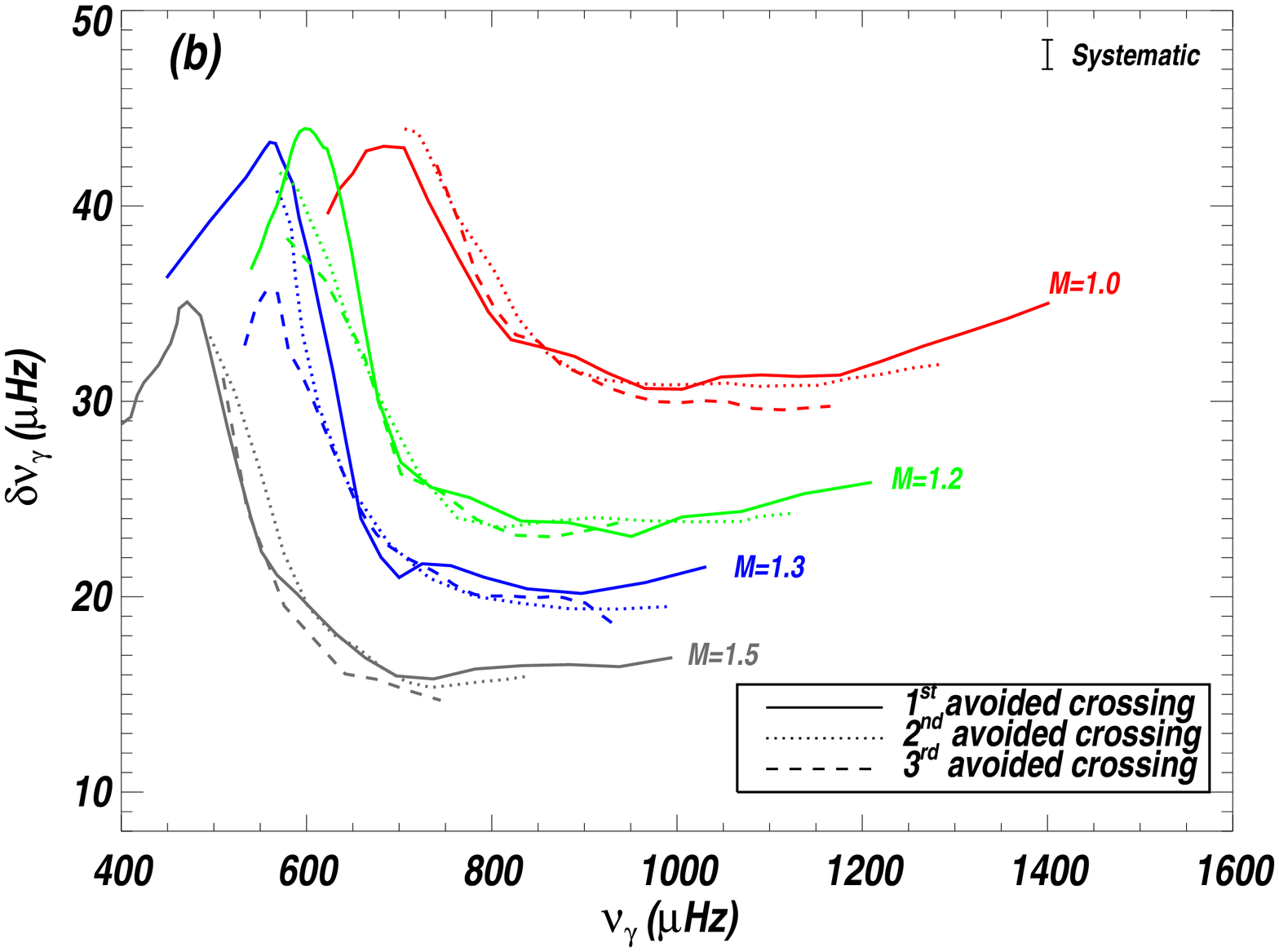} 
\includegraphics*[angle=90,height=6.3cm]{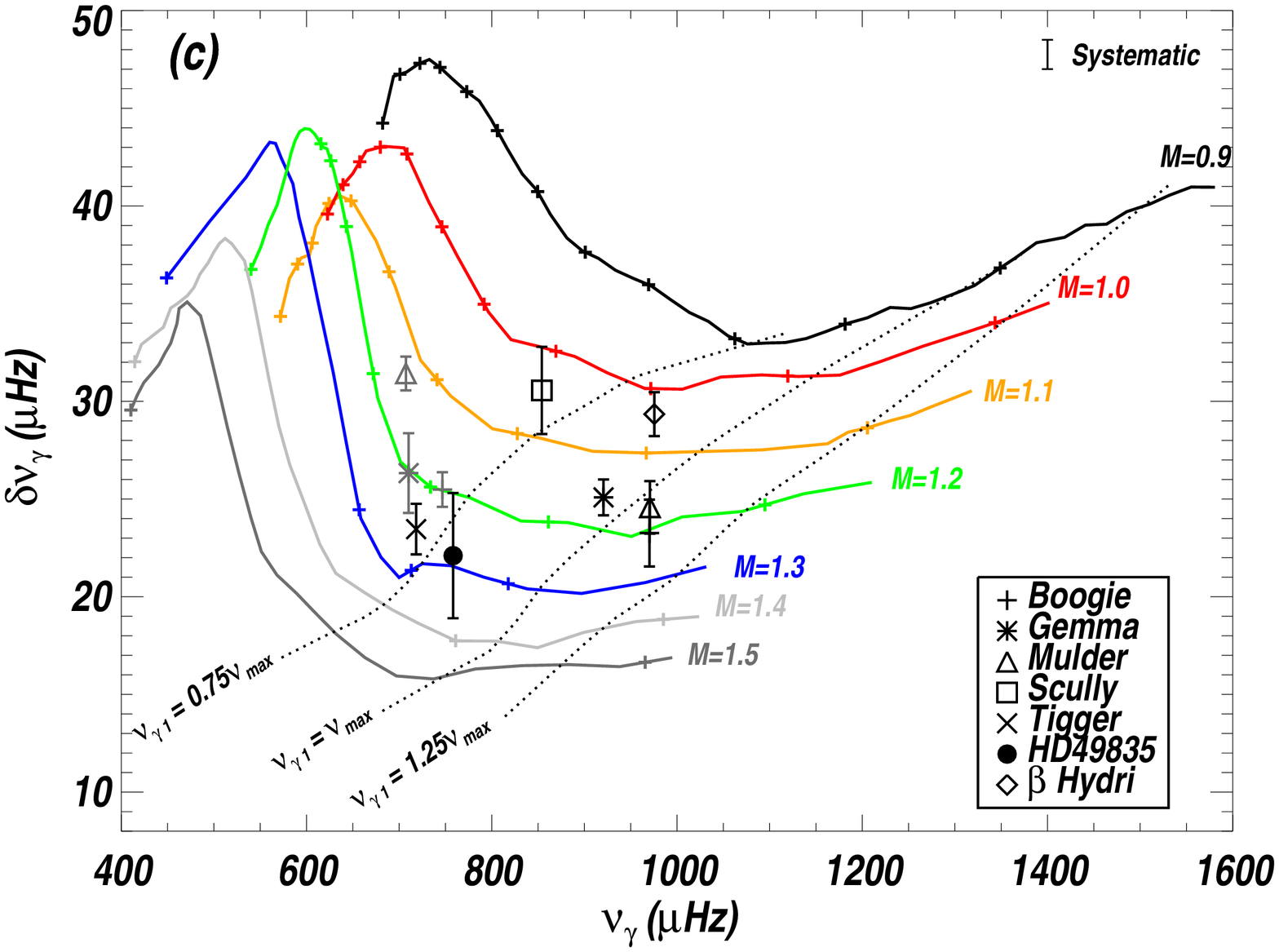} 
\caption{\textbf{(a)}: $\delta\nu_\gamma$ as a function of the
  $\gamma$-mode frequency for $\nu_{\gamma1}$ and for $1.0M_{\odot}$ (red
  lines), $1.2M_{\odot}$ (green lines) and $1.3M_{\odot}$ (blue
  lines). Dashed, solid and dotted lines are for three metallicities Z.%:
  %$Z=0.011$, $Z=0.017$ and $Z=0.028$.% respectively. 
\textbf{(b)}: $\delta\nu_\gamma$ as a function of $\nu_{\gamma1}$ (solid lines),
$\nu_{\gamma2}$ (dotted lines) and $\nu_{\gamma3}$ (dashed lines) at
$Z=0.017$. 
\textbf{(c)}: $\delta\nu_\gamma$ as a function of $\nu_{\gamma1}$ for stellar masses
$M_\odot = 0.9$ to $M_\odot =1.5$, at $Z=0.017$. Crosses are spaced by $150$ Myr. Symbols correspond to the seven
analyzed stars. 
Whenever reliable, fit values for the second measured
avoided crossing are shown (gray symbols). Doted lines indicate when
$\nu_{\gamma1}=0.75\nu_{\rm max}$, $\nu_{\gamma1}=\nu_{\rm max}$ and
$\nu_{\gamma1}=1.25\nu_{\rm max}$. They indicate the best
region to observe the first avoided crossing.} \label{fig:coupling} 
\end{center}
\end{figure}

Figure~\ref{fig:coupling} shows the results, with each panel showing the
coupling strength as a function of the avoided crossing frequency.
The central frequency of the modes $\nu_{max}$ decreases over time and thus, goes from right to left  along a star's evolution in these diagrams. The systematic error on the determination
of $\delta\nu_\gamma$ is about $1.5\,\mu$Hz.  This uncertainty arises from
Eq.\,\ref{eq:LS} not being a perfect representation of the behavior of the
frequencies of the stellar models. In Figure\,\ref{fig:coupling}a we show only the highest-frequency avoided
crossing.  The different tracks show models with four different masses and
three different metallicities.  Interestingly, the coupling strength
depends strongly on mass (as noted by \citealt{Deheuvels2011}) but only
 weakly on metallicity, if at all. The systematic error prevents us from assessing whether the small variations of the coupling strength are really due to the metallicity.
it suggests that we may be able to use the coupling strength to break the
degeneracy between mass and metallicity, which is often a problem in
asteroseismology.

%Figure\,\ref{fig:coupling:gmodefreq} 

As mentioned above, $\gamma$~modes frequencies increase as a
star evolves and several avoided crossing may enter the range of observable
frequencies.  In Figure\,\ref{fig:coupling}b we show $\delta\nu_\gamma$ for the
first three avoided crossings for evolving models with four different
masses.  They all have solar metallicity.  %It is clear that 
 For a given mass, all three curves closely follow the same path in the diagram
(but with a time delay).  This confirms that $\delta\nu_\gamma$, as defined
by Eq.\,\ref{eq:delta_nu_gamma}, is an excellent measure of the coupling
strength between the p- and g-mode cavities.

%The coupling strengths of the avoided crossings depend mainly on the
%frequency of the $\gamma$-modes and not on the evolutionary stage of the
%star, for frequencies higher than $\approx 650\mu$Hz. 

In Figure\,\ref{fig:coupling}{a} and \ref{fig:coupling}{b} we see rapid changes in the
coupling strength at the low-frequency end.  This reflects rapid
changes in the extent of the evanescent zone that separates the inner g-mode and outer p-mode cavities.  Indeed, the coupling
strength depends on the extension of this zone, in the sense that a smaller
evanescent zone leads to stronger coupling.  In Figure\,\ref{fig:coupling}{b} the tracks for the different avoided crossings (for a given mass) no longer follow a single path at low frequency. During the transition between the main sequence and the post main sequence stage, the density of the star and thus, the shape of the evanescent zone, vary on very short timescale. Thus the evanescent zone has changed significantly between the times when the first and third avoided crossing pass
the same frequency.  For the most evolved stars, which means when several $\gamma$ modes have a frequency close to $\nu_{max}$, the diagram is also problematic, as the low-frequency region of the spectrum is densely populated with
$\gamma$ modes, making it hard to obtain a stable fit.

To summarize, we conclude that the coupling strength of mixed $\ell=1$
modes, as measured by the minimal separation $\delta\nu_\gamma$, is a useful
observable in subgiant stars, provided they are not too evolved.  In
particular, $\delta\nu_\gamma$ depends mainly on the stellar mass and is
almost independent of metallicity and of which avoided crossing is being
measured.  %In the next section we 
We will now use this result to estimate the masses of
subgiants for which observed frequencies have been published.

\section{Application to observed frequencies} \label{sec:obs}

We have applied the approach described above to the seven subgiants listed
in Table~\ref{tab:parameters}.  We use the published frequencies, which
were measured either by ground-based spectroscopy, or from space by the
CoRoT or {\em Kepler} missions.  
%To took the uncertainties $\sigma_n$ in the observed frequencies into account, the likelihood was modified to be
%\begin{equation}
%	\ln [ L(\nu_{\rm obs}|\mbox{\boldmath$A$}) ] = - \half \sum_n
%	(\nu_{\rm calc}(n) - \nu_{\rm obs}(n))^2/\sigma_n^2,
%\end{equation} 
%while the priors were unchanged.

Figure\,\ref{fig:coupling}c shows the results. The symbols show the results of applying
our method to the seven subgiants. The colored tracks are from the
analysis of theoretical models discussed above, for a wide range of masses
and a single metallicity, and only for the highest-frequency avoided
crossing (the $\gamma_1$ mode). Crosses along each curve are spaced by 150 Myrs. We see that the $\gamma$-mode frequencies may vary by several hundreds of $\mu$Hz within this time scale while the precision of measure of these frequencies is only of few $\mu$Hz. Thus $\gamma$-mode frequencies provide a stringent constraint on stellar age.

We note that any potential deviation between the tracks of different
metallicity (Figure\,\ref{fig:coupling}a) and for different avoided crossings
(Figure\,\ref{fig:coupling}b) are within the uncertainties with which we can
measure $\delta\nu_\gamma$. We can therefore superimpose results from more
than one avoided crossing per star on the same diagram
(Figure\,\ref{fig:coupling}c).  The quality of the data allowed us to extract
accurately two avoided crossing in three Kepler stars (Mulder, Boogie and
Gemma).

\begin{figure}
\begin{center}
	\subfigure{\epsfig{figure=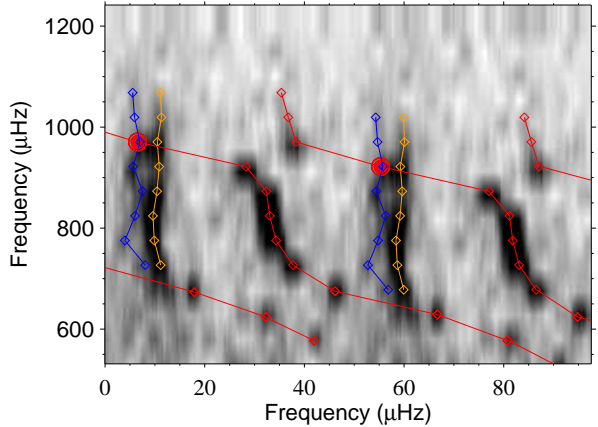,height=8.5cm, angle=90}}\quad
\end{center}
\caption{Replicated \'echelle diagram \citep{bedding2011b} for  Mulder {\em Kepler\/} target).  Two avoided crossings are visible.  The
    greyscale shows the observed power spectrum and open symbols show
    the frequencies reported by \citet{Campante2011}.  The model predicts
    an extra mixed mode (circles), not found by
    \cite{Campante2011} because it lies on an $\ell=2$ mode. }
\label{fig:ech_diag_mulder}
\end{figure}

An example of a fit is shown in Figure\,\ref{fig:ech_diag_mulder} for Mulder.
Despite its simplicity, the linear system of Eq.\,\ref{eq:LS} can be useful
to predict the position of previously unidentified mixed modes. Indeed, our
method predicts an extra $\ell=1$ mixed mode, hardly seen on the power
spectrum because it lies close to an $\ell=2$ mode (see
\citealt{bedding2011b} for another example).  This demonstrates that our
method would allow us to identify $\ell=1$ mixed modes in complicated power
spectra.

The results of the fit are given in Table~\ref{tab:parameters}.  The first
three columns give %the number and name of each star and the observed value
%of $\nu_{\rm max}$, 
the identification number, the name and the observed $\nu_{\rm max}$ of each star. %which is the central frequency of the oscillation power
%excess.  
Columns 4--7 give the fitted values of $\nu_\gamma$ and
$\delta\nu_\gamma$ for the avoided crossings.  The last two columns give
our inferred stellar mass, and the published mass
found using conventional asteroseismic analysis.  For the stars with
published masses, our results agree well. However, in the case of HD~49835, the precision on the determination of $\delta\nu_\gamma$ is low (and so for the mass) because the avoided crossing occurs right at the lower boundary of the observed modes, with only one $\ell=1$ mode below $\nu_{\gamma_1}$. %at
%a frequency that is below the observed modes.

%The uncertainties on the mass of the seven stars are computed using the
%systematic error and the uncertainty on our measure. The latter is
%obtained by linear interpolation between the tracks of different masses
%(cf. Figure\,\ref{fig:coupling}c).

The uncertainties on our mass estimates were obtained by linear
interpolation between the tracks of different masses
(cf. Figure\,\ref{fig:coupling}c).  As already noticed by \cite{Deheuvels2011}
mass and coupling are inversely proportional. Thus, uncertainties increase
with mass because high-mass tracks are closer together than
low-mass tracks.  In the best cases, the total uncertainty (quadratic sum
of the uncertainty of the measurement and of the systematic error) is a few
percent, and in other cases it may be as large as~$10\%$.

The results presented here are based on observed $\ell=1$ mixed
modes. However, the mean large separation computed from the $\ell=0$
frequencies provides an independent way to measure stellar masses.  Thus,
the global accuracy, %for a given star, 
 can be increased by using inferred
masses from both the coupling strength and the mean large separation. %(for a given set of model physics).

With observational data, not all the mixed modes can be seen because the
mode envelope has a limited width.  In order to indicate this in
Figure\,\ref{fig:coupling}c, the upward-sloping curves show where
$\nu_{\gamma_1}=0.75\nu_{\rm max}$, $\nu_{\gamma_1}=\nu_{\rm max}$ and
$\nu_{\gamma_1}=1.25\nu_{\rm max}$.  Thus, these curves show the most likely
region in which to observe the first avoided crossing.  The most likely
region to observe the second and third avoided crossings can be obtained by
shifting these lines towards lower frequencies by $\approx 200 \mu$Hz and
$\approx 400\mu$Hz, respectively.

%			Boogie 					Gemma							John 						Klass						Mulder 					Scully					Tigger				HD49							beta Hydri
%			11395018				11026764					7976303					10018963				10273246				10920273				11234888
%      1.2078150       1.1749784       1.0658649       1.2154981       1.1558607       1.0376897       1.2692612       1.2989234       1.0215879
%     0.068430852     0.056792841     0.052449770     0.075484221     0.048882624     0.055473190      0.10680754      0.12372907     0.070543704

\begin{deluxetable*}{lcrcccccc}\tabletypesize{\footnotesize}
%\rotate
\tablecolumns{9}
\tablewidth{0pc}
\tablecaption{Measured mixed modes properties (minimal separation
  $\delta\nu_\gamma$ and $\gamma$-mode frequency, $\nu_\gamma$) and
  inferred mass $M$ (followed by its uncertainty and systematic error). For
  comparison, M$_{\rm lit}$ lists published masses derived by traditional
  seismic model fitting.} 
\tablehead{
\colhead{Id. Number} & 
\colhead{Nickname} & 
\colhead{$\nu_{\rm max}$ ($\mu$Hz)} &
\colhead{$\nu_{\gamma_1}$ ($\mu$Hz)}  & 
\colhead{$\delta\nu_{\gamma_1}$ ($\mu$Hz)} & 
\colhead{$\nu_{\gamma_2}$ ($\mu$Hz)} &
\colhead{$\delta\nu_{\gamma_2}$ ($\mu$Hz)}  &   
\colhead{M (M$_\odot$) } &
\colhead{M$_{\rm lit}$ (M$_\odot$)} }
\startdata
KIC 10273246 \tablenotemark{(a)}   & Mulder   & $842$ & $970.7 \pm 3.2$  & $24.6 \pm 1.4$   & $707.3 \pm 1.2$ & $31.4 \pm 0.9$ &  $1.16 \pm 0.03$ $(\pm 0.04)$  \\
KIC 10920273 \tablenotemark{(a)}  & Scully   &  $974$ & $854.1 \pm 2.6$  & $30.6 \pm 2.3$  &                 &                 &  $1.04 \pm 0.04$ $(\pm 0.04)$ \\
KIC 11234888 \tablenotemark{(b)}  & Tigger   &  $675$ & $718.4 \pm 5.6$  & $23.5 \pm 1.3$  &                 &                 &  $1.27 \pm 0.03$ $(\pm 0.10)$ & $1.33 \pm 0.26$  \\
KIC 11395018 \tablenotemark{(b)}  & Boogie   &  $830$ & $970.2 \pm 10.6$ & $23.3 \pm 1.7$  & $747.7 \pm 0.6$  & $25.6 \pm 0.9$ &  $1.21 \pm 0.06$ $(\pm 0.04)$ & $1.25 \pm 0.24$ \\
KIC 11026764 \tablenotemark{(c)} & Gemma    & $857$  & $920.7 \pm 3.1$  & $25.1 \pm 1.0$  & $710.2 \pm 5.3$  & $26.3 \pm 2.1$  &  $1.17 \pm 0.03$ $(\pm 0.05)$ & $\simeq 1.13$ or $1.23$  \\
HD 49835   \tablenotemark{(d)}    &         &  $1013$ & $758.3 \pm 3.2$  & $22.1 \pm 3.2$  &                 &                  &  $1.30 \pm 0.11$ $(\pm 0.06)$ &  $1.25 \pm 0.05$  \\
	$\beta$ Hydri \tablenotemark{(e)} &         & $960$ & $975.5 \pm 1.6$  & $29.3 \pm 1.2$   &                &                   &  $1.02 \pm 0.06$ $(\pm 0.04)$ &  $1.08 \pm 0.03$
\enddata \label{tab:parameters}
\tablenotetext{(a)}{\cite{Campante2011}} \tablenotetext{(b)}{\cite{Mathur2011}}  \tablenotetext{(c)}{\cite{Metcalfe2010}} \tablenotetext{(d)}{\cite{Deheuvels2010a} and \cite{Deheuvels2011}} \tablenotetext{(e)}{\cite{Brandao2011}} 
\end{deluxetable*}

%\section{Conclusion and perspectives} \label{sec:conclusion}
%
%In subgiant stars, mixed modes arise from a coupling between $p$-modes in
%the envelope with $g$-modes in the core. Here, we presented an approach to
%measure the main characteristics of this coupling. We applied our method to
%frequencies of stellar evolution models and we showed that the coupling
%strength depends mainly on the stellar mass, while being relatively
%insensitive to metallicity or to which avoided crossing is being
%measured. The approach was then applied to seven subgiant stars, allowing
%us to determine their mass.

Theoretically, the coupling strength should depend on the extent of the
evanescent zone between the $g$-mode cavity in the core and the $p$-mode
cavity of the envelope. Thus, it would be interesting to understand the
exact nature of this relation, and why the minimal separation $\delta\nu_\gamma$ seems to
depend strongly on stellar mass whereas it is the same for all avoided
crossings and for a given star.

In this Letter, the effects of varying the mixing length parameter and the initial helium abundance have not been tested. They are known to be correlated to stellar mass and deserve further investigation.

\clearpage

\bibliographystyle{apj}
%\bibliography{ma_biblio}

\clearpage

%% the file, uncomment the \plotone, \plottwo, and
%% If you need a layout that cannot be achieved with \plotone or
%% \plottwo, you can invoke the graphicx package directly with the
%% \includegraphics command or use \plotfiddle. For more information,
%% please see the tutorial on "Using Electronic Art with AASTeX" in the
%% documentation section at the AASTeX Web site,
%% http://www.journals.uchicago.edu/AAS/AASTeX.
%%
%% The examples below also include sample markup for submission of
%% supplemental electronic materials. As always, be sure to check
%% the instructions to authors for the journal you are submitting to
%% for specific submissions guidelines as they vary from

%% This example uses \plotone to include an EPS file scaled to
%% 80% of its natural size with \epsscale. Its caption
%% has been written to indicate that additional figure parts will be
%% available in the electronic journal.

\clearpage

\clearpage

\end{document}